# Network analysis of ballast-mediated species transfer reveals important introduction and dispersal patterns in the Arctic

Mandana Saebi[1,2,+], Jian Xu[1,3,+], Salvatore R. Curasi[4,+], Erin K. Grey[5], Nitesh V. Chawla[1,2,] David M. Lodge[6*]

[1]Department of Computer Science and Engineering, University of Notre Dame

[2]Interdisciplinary Center for Network Science and Engineering (iCeNSA)

[3] Citadel LLC,

[4]Department of Biological Sciences, University of Notre Dame

[5]Division of Science, Mathematics and Technology, Governors State University

[6]Cornell Atkinson Center for Sustainability, and Department of Ecology and Evolutionary Biology, Cornell University

+ The authors contributed equally to the manuscript.

**Abstract.** Rapid climate change has wide-ranging implications for the Arctic region, including sea ice loss, increased geopolitical attention, and expanding economic activity, including a dramatic increase in shipping activity. As a result, the risk of harmful non-native marine species being introduced into this critical region will increase unless policy and management steps are implemented in response. Using big data about shipping, ecoregions, and environmental conditions, we leverage network analysis and data mining techniques to assess, visualize, and project ballast water-mediated species introductions into the Arctic and dispersal of non-native species within the Arctic. We first identify high-risk connections between the Arctic and non-Arctic ports that could be sources of non-native species over 15 years (1997-2012) and observe the emergence of shipping hubs in the Arctic where the cumulative risk of non-native species introduction is increasing. We then consider how environmental conditions can constrain this Arctic introduction network for species with different physiological limits, thus providing a species-level tool for decision-makers. Next, we focus on within-Arctic ballast-mediated species dispersal where we use higher-order network analysis to identify critical shipping routes that may facilitate species dispersal within the Arctic. The risk assessment and projection framework we propose could inform risk-based assessment and management of ship-borne invasive species in the Arctic.

**Introduction**

Global trade and transportation networks can introduce non-native species to ecosystems via land, sea, and air as evidenced by thousands of reported case [1]–[4]. As trade and transportation volumes and connectivity increase, so does the potential for human-assisted species introduction via these mechanisms [4]–[8]. When introduced to a compatible environment, a species can become established, with a subset of established species becoming invasive, i.e., threatening to the economy, environment or human health[9]. The World Wildlife Fund estimates that between 2004 and 2009, aquatic invasive species caused at least 50 billion dollars of damage to fisheries, aquaculture, water supply systems, industrial infrastructure and harbors [10]. Established invasive species are in many cases impossible or expensive to eradicate, and eradication efforts can harm native species [10],[11]. The difficulty of eradication coupled with the potential economic impact of invasive species underscores the high value of prevention [9]. Improved management of invasive species, including prevention strategies, is a primary goal of multiple international agreements including the Convention on Biological Diversity. A detailed understanding of species introduction risk through transportation networks [10],[11] is an important prerequisite for effective prevention.

The global shipping network is the dominant vector for the unintentional introduction of aquatic species to new ecosystems [12], typically through ballast water discharges and biofouling (i.e., organisms attached to the surfaces of ships) [6],[11],[13]–[15]. Prerequisites for ship-borne species invasion include that the species survive transportation, establish in the new environment, and spread [6],[11],[16]. These ship-borne species introduction processes are in turn influenced by multiple factors including: ship type; voyage duration; ballast water uptake volume and location, discharge volume and location; environmental differences between the source and destination ports, and the environmental tolerance of the organisms [6][16]–[18]. Therefore, assessing species introduction and dispersal risks is a complex task involving many types of data from many different sources.

Aquatic invasive species are a widely-recognized threat to Arctic ecosystems [14]. The complex interplay between climate change, shipping activities, and environmental conditions in the Arctic adds dimensionality to the risk evaluation [11],[14],[16]. In particular, climate change is decreasing Arctic sea ice extent, leading to increased shipping, changes in shipping patterns, and increased human activity and interest in the Arctic [19]. In recent years the number of ships traveling through the Arctic has increased 20% annually, leading to an associated increase in species introduction risk [20]. Climate change may expand the range of invasive species by altering the temperature and salinity of ports, allowing species to survive in locations they could not previously [18]. Here we consider how current shipping trends and climate interact to influence non-native species introduction into and dispersal within the Arctic.

The complex process of ship mediated species introduction, the economic importance of global shipping (~90% of global trade [21]), the interactions between stakeholders and nations within the Arctic, and the uncertain effects of climate change mean that addressing the issue of aquatic invasive species in the Arctic is complex, important and urgent [9],[16]. In the May 2017 Fairbanks Declaration [22], the eight member countries of the Arctic Council endorsed an action plan to reduce the impact of invasive species in the Arctic, emphasizing the importance of shipping as a primary pathway of species introductions [23].

Here we employ network analysis and data mining techniques to assess, visualize, and project aquatic species introduction into and dispersal within the Arctic via shipping. The network approach we use does not substitute existing analysis based on pairwise species introduction risks between two ports [6]. Rather, it provides a unique perspective on how species introduction affects multiple ports in a tightly coupled cluster, and how a ships' previous locations can dictate subsequent species introduction risks in the Arctic. Without such analyses, it will be impossible to prioritize surveillance, prevention, and other management efforts among ports and routes, which will be necessary to achieve the goals laid out by the Arctic Council [23].

We accomplish these goals by building a risk assessment network model tailored for shipping into and within the Arctic. We use the best available global data sets including ship movement data [24] from Lloyd's List Intelligence, an Informa Group Company (LLI, New York, NY, USA). LLI collects global port calls in conjunction with the Lloyd's Agency Network, which gathers information from over 1200 local agents who observe port arrivals and departures directly and additionally gather data from other sources. The LLI Database is a global standard for commercial ship traffic research for government and business applications. We also collect ballast water discharge data [25], biogeographical data [26],[27], and environmental data (temperature, salinity) [24],[28]. Building on previous modeling frameworks to assess the relative risk of introduction posed by ships [6],[16],[29]–[31], we leverage the network approach in [31] for evaluating the current and future relative risk of species introduction into and dispersal within the Arctic posed by ballast water discharges, which could readily be extended to risks posed by biofouling [16]. We use existing models and metrics for individual components of our analyses such as ballast water discharge models or environmental tolerance metrics. Our focus is on integrating multiple factors that impact species invasion into a unified network analysis framework to model species invasion risk.

Our analysis includes the following components. First, we use all voyages that originate outside the Arctic and end in the Arctic to estimate the relative risk of species introduction based on shipping frequency, ship size, ship type, trip duration, and ballast water exchange patterns in a first-order network. Under the assumption that current climate-change driven shipping trends continue, we analyze and project the evolution of the network, and illustrate the emergence of shipping hubs in the Arctic. Second, we investigate the influence of species' sensitivity to environmental differences between their origin ports outside the Arctic and destination ports inside the Arctic by visualizing how the topology of the species introduction network adapts to environmental constraints on species establishment.

Third, we shift our attention from identifying the Arctic ports at highest risk of initial introduction and establishment to the relative risk of subsequent ballast-mediated dispersal among ports within the Arctic. We compare intra-Arctic species dispersal risk estimated using a first-order network to that estimated using a higher-order network [30],[31] which incorporates the dependency of a ship's next destination on the origin of its previous voyage. Finally, we use a case study to demonstrate how the higher-order network can help identify high risk species dispersal pathways and inform the development of more targeted management policies.

Our analyses are novel and important for a number of reasons: (1) this is the first risk assessment network model tailored for species introduction into and dispersal within the Arctic; (2) we demonstrate the recent emergence of shipping hubs in the Arctic and discuss the implications for the control of species invasion; (3) we offer managers an alternative to generic port-pair spread risks by breaking down species into environmental tolerance groups and predicting their respective introduction pathways; (4) we illustrate how higher-order network modeling can provide more realistic risk estimates and can help narrow down potential species dispersal pathways for management. Our framework and results provide a foundation for risk-based prioritization of surveillance among Arctic ports and for efforts to prevent non-native species invasions in the Arctic.

**Results**

**Species introduction to the Arctic**

In the global shipping network comprised of nodes (representing ports) and edges (representing direct shipping routes among ports), we first investigate species *introduction pathways* to the Arctic (direct ship movements connecting Arctic ports to non-Arctic ports, illustrated in Fig. 1A as the purple lines connecting the green non-Arctic ports to orange Arctic ports). The introduction pathways naturally form a *bipartite network*: on one side are the 310 Arctic ports (using the Arctic conservation area boundary defined by the Arctic Council), and on the other side are the 7,187 non-Arctic ports that could be the direct source of invasive species. From 1997 to 2012, voyages in the LLI data form 3,902 active introduction pathways. We exclude pathways from the same or neighboring ecoregions (geographic regions with similar environmental conditions), because species could possibly disperse to these nearby locations naturally [26],[27]. Removing pathways using these ecoregion criteria further reduces the number of introduction pathways we analyzed to 2,874.

With increased human activities in the Arctic, the properties of the species introduction network evolved over time (Fig. 1B, Supplementary Table S1) from 1997 to 2012. We use a simple linear model because we have no strong basis for expecting any particular alternative functional form, and the linear model makes minimal assumptions. Linear trend lines, which we do not convey as predictions, plus confidence intervals are projected for the next 15 years. In general, we observe an increase in shipping activities in the Arctic from 1997 to 2012, shown by the significant increase of number of voyages (+128/year, $p < 0.05$), the total dead weight tonnage (DWT, a crude estimate of propagule pressure) of the ships (+2.52×$10^{16}$/year, $p < 0.01$), and the average capacity of ships (DWT per voyage, +253/year, $p < 0.05$). On the other hand, the number of distinct introduction pathways only increased by 2% over the same period ($p > 0.05$). These observations indicate that the increased shipping activities in the Arctic are reflected in heavier shipping traffic per introduction pathway, as opposed to an increase in the number of introduction pathways.

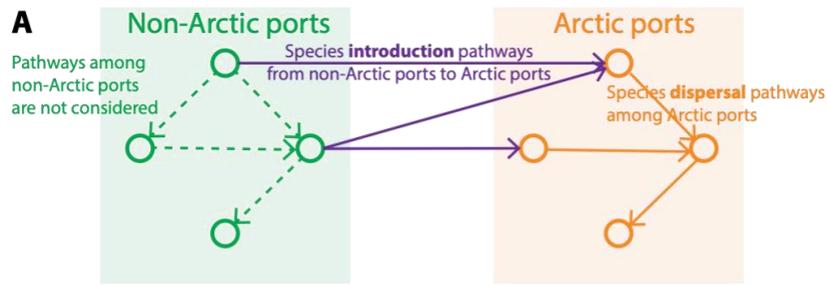
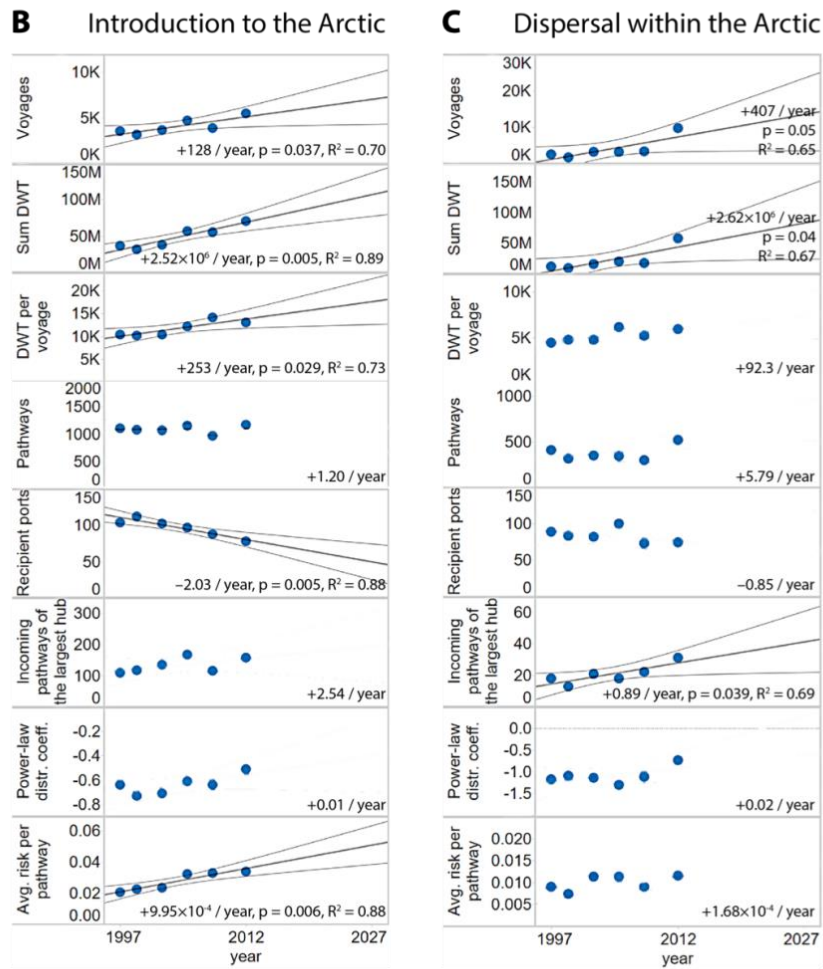

**Fig. 1: The evolution of shipping activities and species introduction risks in the Arctic. (A) Diagram illustrating** species introduction pathways (from non-Arctic port to Arctic port) and dispersal pathways (from Arctic port to Arctic port). **(B)** The evolution of species introduction pathways from non-Arctic ports to Arctic ports. **(C)** The evolution of species dispersal pathways within the Arctic. 95% confidence intervals for projections are given for regressions that have $p < 0.05$.

We further observe the emergence of shipping hubs in the Arctic. Counter-intuitively, the number of recipient ports per introduction pathways has been significantly decreasing (-2.03/year, $p < 0.01$) despite increased shipping activity in the Arctic (Fig. 1B). As a result, a few ports have experienced a significant increase in introduction pathways (Murmansk's introduction pathways grew from 110 in 1997 to 158 in 2012). The evolution of the species introduction network shows that the shipping traffic was previously more evenly distributed amongst Arctic ports but is gradually being "rewired" to a few hub ports in the Arctic. From the networks perspective, this rewiring process is a real-world example of the network evolution following *preferential attachment [32]*. The relationship between the number of incoming pathways $p$ and the number of ports with $p$ incoming pathways $f_p$ (called the *degree distribution* in network science and graph theory [32]) follows a power-law distribution $f_p = ap^b$ [33]. We observed that the power-law coefficient $b$ (computed using Adam Ginsburg's Python package) increased from -0.64 to -0.51 over the 15 years, indicating a longer tailed distribution and the emergence of highly connected hubs (Fig. 1B).

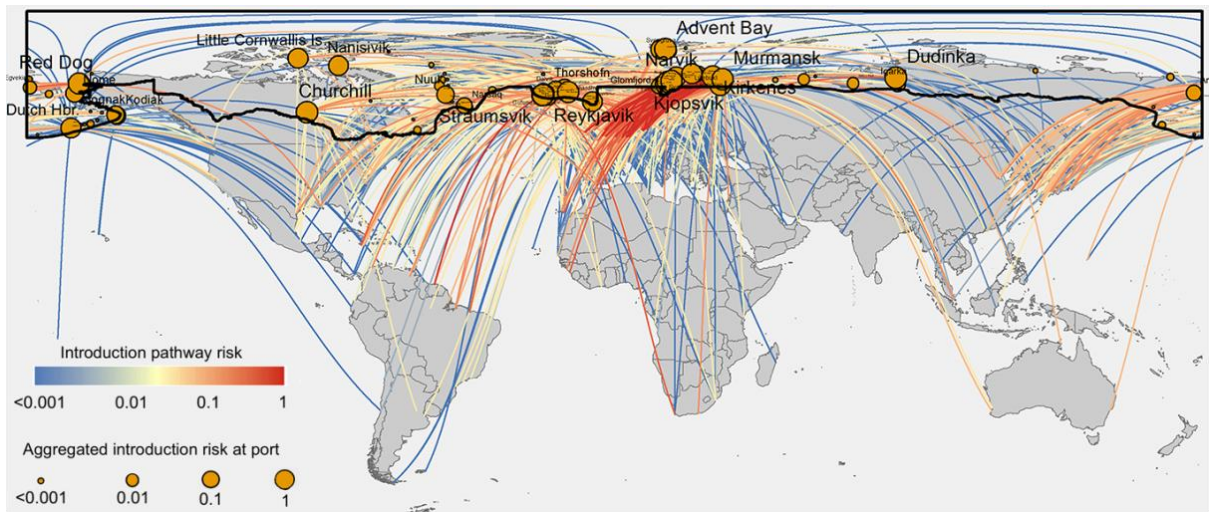

**Fig. 2: A global overview of species introduction pathways into the Arctic via shipping**. Colors of links indicate the relative risk of introduction $P_{i \rightarrow j}$ from non-Arctic port $i$ to Arctic port $j$. Sizes of nodes indicate the risk of introduction to that port $P_j$ aggregated over all voyages into that port. Relative introduction risks and port introduction risks calculated using shipping data from 1997-2012. The black outline delineates the Arctic boundary.

Focusing on individual introduction pathways, we denote the *relative risk of species introduction* from a non-Arctic port $i$ to Arctic port $j$ as $P_{i \to j}$. Note that $P_{i \to j}$ is not an absolute probability of species introduction, but a robust metric of the relative risk of species introduction determined by shipping frequency, ship size, ship type, trip duration, ballast water exchange amount and frequency (see Methods section; environmental similarities that influences risk of establishment will be discussed in-depth in the next section). In general, over the 15 years, we observe a significant increase in the average $P_{i \to j}$ per introduction pathway (+9.99×10$^{-4}$/year, p < 0.01), a combined effect of the increase of shipping frequency, ship size, and the relatively stable number of introduction pathways (Fig. 1B).

We visualize the species introduction pathways in Fig. 2, with colors indicating relative risk of introduction $P_{i \to j}$ across 15 years from 1997 to 2012. Many high-risk pathways (red paths in Fig. 2) originate from Northwestern Europe (e.g., Rotterdam, Hamburg, and Amsterdam) and point to Arctic ports of Narvik in Norway and Murmansk in Russia (Supplementary Table S2). For each port in the Arctic we aggregate the risk of introduction denoted by node sizes in Fig. 2. Ports associated with high-risk pathways (such as Murmansk) demonstrate high aggregated risks; however, even some ports such as Afognak and Kodiak (both in Alaska) associated with low-risk pathways demonstrate high aggregated risks because of the substantial number of different ports to which they are connected. Full data of species introduction pathways (per year and overall) are available online.

**Risk of species establishment in the Arctic**

We now consider the environmental constraints on species establishment, and how these will alter the topology of the species introduction network. Fig. 2 shows that many introduction pathways connect the Arctic to distant ports in Australia, South America, and Africa. Many species are unlikely to survive translocation between these ports and the Arctic because of the temperature and/or salinity differences between these regions. Similarly, species being transported from the freshwater Great Lakes ports would likely not establish in a marine Arctic port due to wide salinity differences. To present a more detailed view of establishment risk to the Arctic, we categorize species into six groups that reflect different environmental tolerances to temperature changes (Δt

≤ 2.9° and Δt ≤ 9.7°) and salinity changes (Δs ≤ 0.2ppt, Δs ≤ 2ppt, and Δs ≤ 12ppt) [16], based on estimated long-term thermal tolerances of marine invertebrate taxa [34]. This analysis illustrates which Arctic species introduction pathways are most likely to lead to species establishment for species in a given tolerance group (Fig. 3, Supplementary Table S3).

We first looked at the introduction pathways connecting Arctic ports with non-Arctic ports that have almost identical environmental conditions (Fig. 3 top left, Δt ≤ 2.9°, Δs ≤ 0.2ppt). Although the number of introduction pathways available to species in this tolerance group are limited, these introduction pathways need particular attention from policy makers, because all species groups (including those most sensitive to environmental changes), will have the potential to establish in the target Arctic port once introduced. These pathways including those from Port Alfred (Canada) to Churchill (Canada), and from Seaham (UK) to Akranes (Iceland).

Our analysis can also inform decision-makers regarding the role of temperature and salinity in shaping the species introduction network. Churchill (Canada) has the highest combined risk of species introduction $P_j = 3.6\%$ (see Methods) for the most sensitive species group (Δt ≤ 2.9°, Δs ≤ 0.2ppt). When the temperature tolerance constraint is loosened to Δt ≤ 9.7° and salinity tolerance held constant (Δs ≤ 0.2ppt), Churchill (Canada) remains the most vulnerable port with $P_j$ unchanged (Fig. 3, top right). But when the salinity constraint is loosed to Δs ≤ 2ppt and temperature tolerance held constant (Δt ≤ 2.9°), Afognak (USA) and Dutch Harbor (USA) emerge as the more vulnerable ports (with $P_j$ = 5.6% and 4.7%, respectively), because species from ports in Northeastern Asia (notably Japan and Russia) can likely survive in the new environment (Fig. 3, mid left). In general, going from top left to bottom right on Fig. 3, we can learn which introduction pathways are open to species that are more resilient to temperature or salinity changes. Such information makes it possible to focus on the physiological limits of organisms when devising targeted management strategies for specific routes. Full data of species introduction routes and aggregated risks at ports under different environmental constraints are available online.

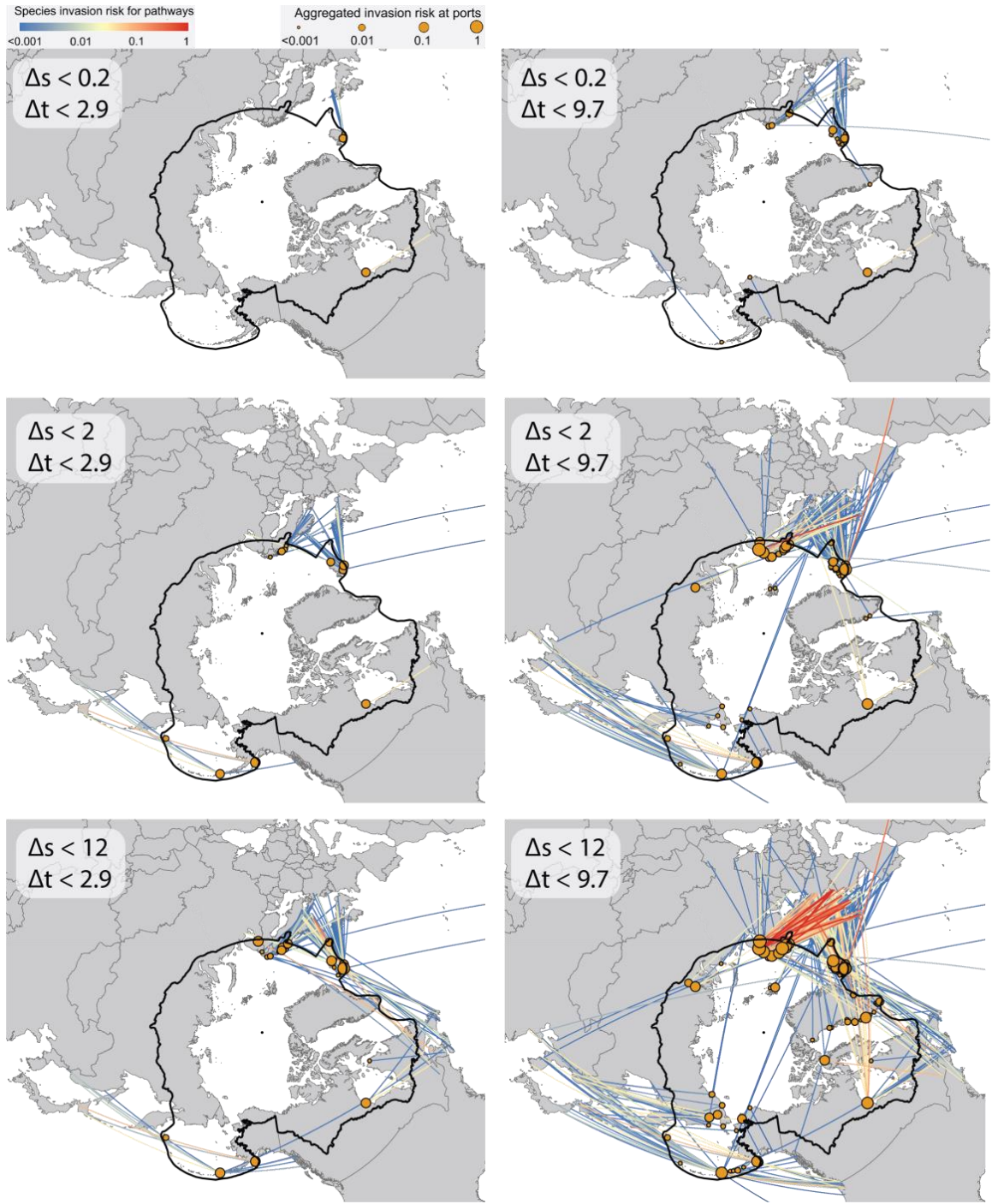

**Fig. 3: The influence of species' sensitivity to environmental change on spread pathways.** Species spread (introduction + establishment) pathways from shipping originating outside the Arctic, organized by environmental tolerance groups, with link colors indicating the relative risk of spread and node size indicating the aggregated spread risk. Aggregated spread risk per node increases when considering species with broad environmental tolerances

**Ship-borne species dispersal within the Arctic: Direct and Indirect pathways**

Here we shift our attention from identifying the Arctic ports at highest risk of initial introduction and establishment to the relative risk of subsequent ship-driven dispersal among ports within the Arctic, referred to as *dispersal pathways* (orange pathways in Fig. 1A). We model species dispersal as a network: nodes represent Arctic ports, links that connect nodes represent the species dispersal pathways, and links' weights (strengths) represent the relative risks of pathways. From 1997 to 2012, voyages in the LLI data form 1,269 direct dispersal pathways within the Arctic (Fig. 4). Most high-risk connections (denoted with thick lines) exist between ports in Arctic Europe, including those connecting Murmansk in Russia and Tromso in Norway (top 10 risky pathways in Supplementary Table S2). The aggregated risks of dispersal (denoted with node sizes) at ports in Norway and Iceland are particularly high, due to the many strong dispersal pathways connecting to them. A projection of the species dispersal within the Arctic is shown in Fig. 1C. We observe a significant increase in the number of voyages, ship size, and the number of incoming pathways from the largest hub (+0.89/year, $p < 0.04$). Full data are available online in our public repository.

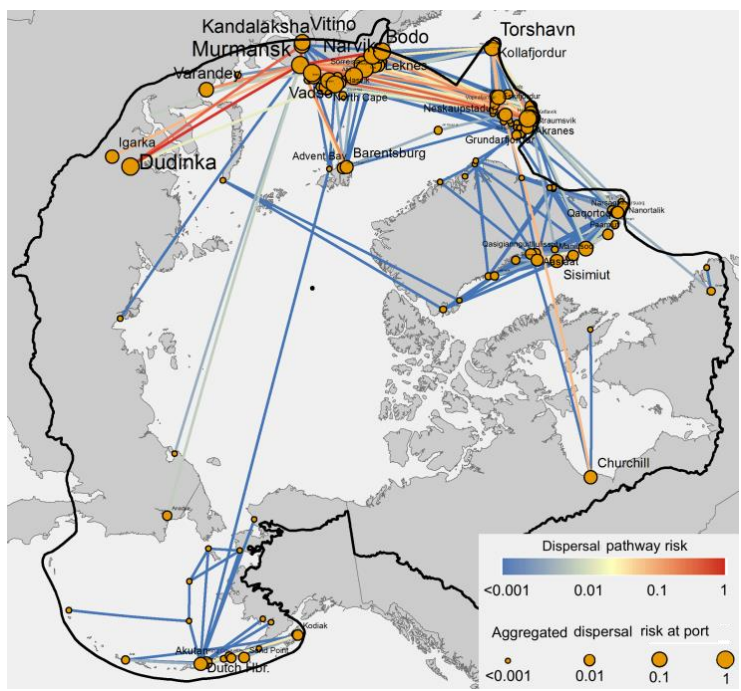

**Fig. 4: Species dispersal pathways within the Arctic.** Colors of links indicate the relative risk of dispersal. Sizes of nodes indicate the aggregated dispersal risks for direct intra-Arctic species dispersal.

The analyses above focus on the *direct* dispersal pathways pointing to a ship's next destination from the ship's current port but ignore the fact that a ship's next destination can also depend on its previously visited ports. Ships' multi-step movement patterns (i.e., *higher-order* movement patterns) can help refine the modeling of *indirect* species dispersal through intermediate ports. In the illustrative example in Fig. 5A we consider six ships coming from Tromso or Narvik, going through Murmansk, and heading to Bodo or Nuuk. The conventional first-order network analysis [16],[35] simply counts the number of trips between port pairs to model direct species dispersal. We refer to this network as first-order network of species flow (SF-FON). In this case, the number of trips, and therefore edge weight or "risk", is three for all trips between ports including from Tromso to Murmansk, from Narvik to Murmansk, from Murmansk to Bodo, and from Murmansk to Nuuk (Fig 5B, left). However, according to the ship trajectories in Fig. 5A, ships coming from Tromso through Murmansk are more likely to go to Bodo (edge weight two) than Nuuk (edge weight one), and ships coming from Narvik are more likely to go to Nuuk than Bodo. If the ballast water management at Murmansk is not 100% effective (i.e., ship did not fully discharge or treat ballast water at Murmansk), species from Tromso are more likely to reach Bodo than Nuuk through the intermediate port Murmansk, following the higher-order ship movement patterns. This important information on indirect species dispersal is completely hidden from SF-FON (Fig. 5B, left), but is obvious if a higher-order network representation [31] is used. Therefore, in this section, we use higher-order network of species-flow (SF-HON)[31]. Fig. 5B, right shows an example. SF-HON splits the node Murmansk into two nodes "Murmansk given the last port being Tromso" (Murmansk|Tromso) and "Murmansk given the last port being Narvik" (Murmansk|Narvik), each node having differently weighted outgoing edges to Bodo and Nuuk. Decision-makers can derive this additional information on indirect species dispersal from SF-HON that they would miss if they relied merely on the SF-FON analysis.

The example of considered SF-HON structure and dynamics outlined above is likely to be very important to species dispersal: ship movements demonstrate up to fifth order dependency which can only be effectively captured by higher-order network modeling [30],[31]. Moreover, higher-order structures influence clustering, ranking, and diffusion in networks [30],[36]. However, SF-HON structure and its implications have not been studied for Arctic shipping. Here we provide a systematic analysis of the Arctic species dispersal network using SF-HON and compare that to Sf-FON. The details of SF-HON construction are provided in the Methods section and supplementary Fig S1).

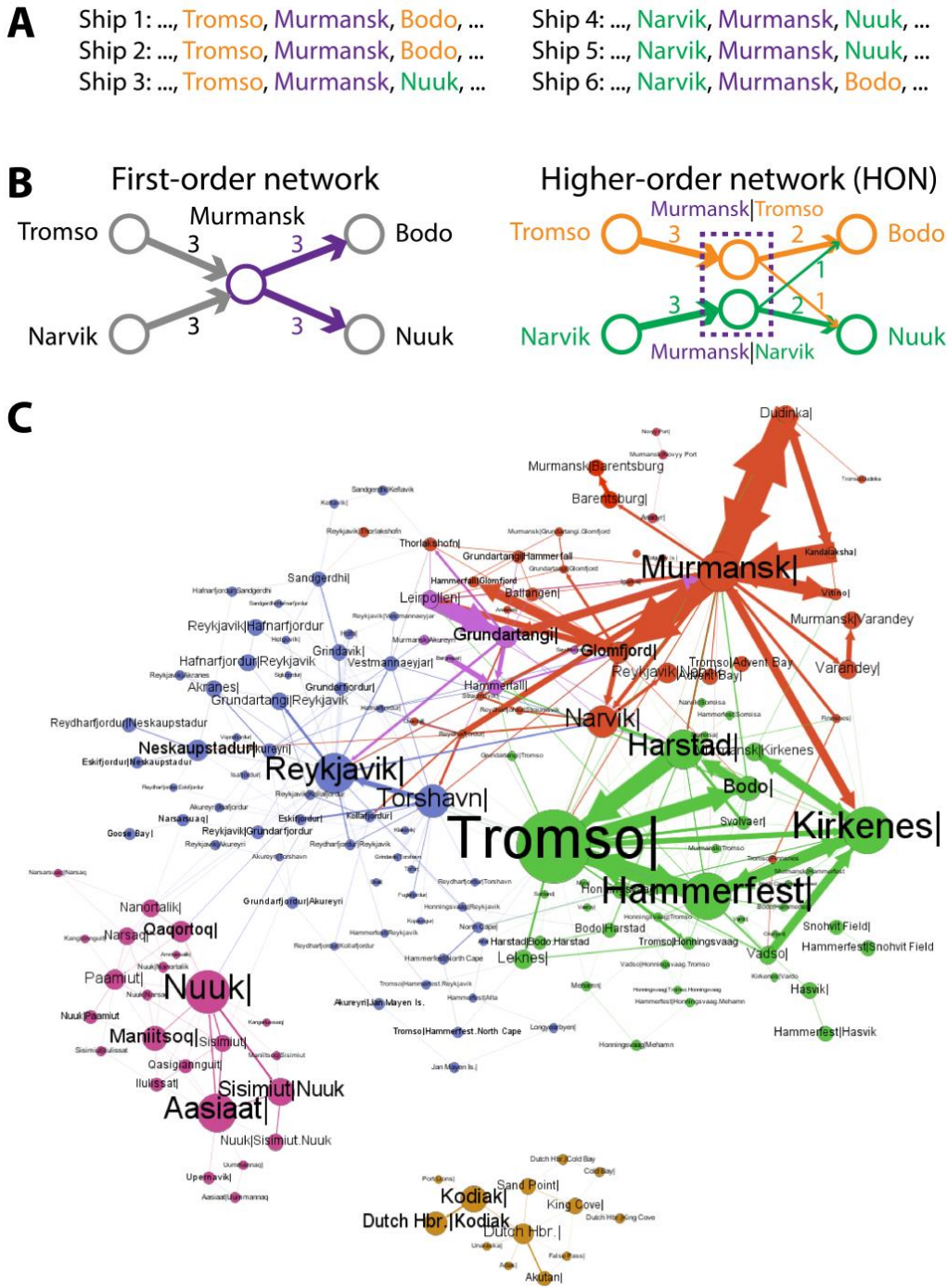

**Fig. 5: Species dispersal higher-order network in the Arctic. (A)** Example ship movement data used to create the network in Fig. 5B **(B)** Example of species dispersal represented as a higher order network which considers that a ships next movement and species dispersal can depend on multiple previous steps. **(C)** Species dispersal higher-order network in the Arctic. Clusters of ports tightly coupled by species dispersal pathways are distinguished by colors. Multiple nodes with the same [CurrentPort] represent the same physical location but with different previous locations. The size of nodes represents the relative probability that species end up at the given port by randomly flowing through the SF-HON in multiple steps.

We first visualize the SF-HON of Arctic species dispersal pathways in Fig. 5C, where nodes represent ports (with labels in the form of [CurrentPort] | [PreviousPorts]), edges (lines connecting nodes) represent species dispersal pathways between Arctic ports, and edge weights (line widths) represent species dispersal probabilities (only non-trivial pathways with $P_{i \to j} \geq 0.001$ are shown). To highlight the connections among ports, instead of placing ports on a map, we use a layout that places nodes with stronger connections closer to each other [37]. Ports in the network are automatically grouped into six distinct *clusters*; connections within each individual cluster are denser and stronger than connections between different clusters. Although the algorithm behind the layout in Fig. 5C does not include any geographical information, ports still self-organize into clusters that are clearly separated geographical regions: specifically, the Greenland cluster (magenta) has only a few weak connections to the Icelandic cluster (blue), and the Alaskan cluster (brown) has no significant dispersal pathways to other clusters. This provides opportunities for more effective species management: species control policies targeting the loose connections between clusters (such as those between Greenland and Iceland) can effectively prevent or slow species propagation from one cluster to another with management target at only a few voyage routes.

We next show how the dynamics of indirect species dispersal in the Arctic is more accurately reflected in SF-HON than in SF-FON. Species dispersal can be thought of as a *random walk* process on the network: starting from a port $i$ in the Arctic, the species will randomly travel through one of the outgoing connections as its next step of dispersal, and the probability of choosing the connection is proportional to the relative strength of the introduction risk $P_{i \to j}$; that is, $\widehat{P_{i \to j}} = \dfrac{P_{i \to j}}{\sum_k P_{i \to k}}$. In the SF-FON model in Fig. 5B left, a random walker (representing a species) starting from Tromso may first disperse to Murmansk through shipping, then have 50/50 chances of dispersing directly from Murmansk to Bodo and Nuuk. In reality, before the species is established in Murmansk and has reached a sufficient propagule pressure to disperse from Murmansk, the species would have already been carried by the same ship to the next port (Bodo/Nuuk), causing indirect dispersal from Tromso to Bodo/Nuuk (assuming the ship's ballast water is not fully treated at Murmansk). This indirect species dispersal behavior is more accurately reflected when random walkers move along the SF-HON model (Fig. 5B right), which will have 2/3 and 1/3 chance dispersing from Tromso through Murmansk to Bodo and Nuuk, respectively. In other words, the species dispersal modeled by random walk on SF-HON is not completely random like on SF-FON but follows higher-order ship movement patterns.

We further show that the relative rankings of ports by species dispersal risk are different under SF-FON and SF-HON network representations. We approximated species dispersal dynamics using the well-established PageRank [38] "random walking with resets" algorithm. This is shown in Fig. 5C, where the node size indicates the port's risk of receiving indirect species. These species dispersal risks are fundamentally different from the aggregated risk of receiving species dispersal illustrated earlier in Fig. 4, which was computed on SF-FON and only aggregates risks of *direct* species dispersal pathways to a port, regardless of ships' higher-order movement patterns. A side-by-side comparison of port risks of direct dispersal on SF-FON and indirect dispersal on SF-HON is provided in Supplementary Table S4. Taking Reykjavik in Iceland as an example, the direct risk of receiving species dispersal for Reykjavik only ranks 14[th] among all Arctic ports, since the pathways connecting to Reykjavik are weak ( $\max_i(P_{i \to \text{Reykjavik}}) = 0.14$ compared to $\max_i(P_{i \to \text{Murmansk}}) = 0.70$ ). However, Reykjavik ranks 3[rd] for receiving indirect species dispersal, and is the central port in the Icelandic cluster, therefore, species at other ports in the Icelandic cluster have a high probability of eventually flowing to the topologically highly connected port of Reykjavik.

The case for Reykjavik as a high-risk recipient port of indirect species dispersal is consistent with recent reports of several ship-borne non-native species establishing themselves in southwest Iceland [39]. Another example supporting the SF-HON species dispersal is the presence of at least 8 cryptogenic or nonindigenous species in Dutch Harbor Alaska [40], a port that ranks 8[th] in the risk of receiving indirect dispersal but only 24[th] in the risk of receiving direct dispersal. Unfortunately, more complete and rigorous testing of SF-FON and SF-HON risk rankings in Arctic ports is not possible given the absence of standardized biological data. Additionally, since new species may take years or decades to establish and become detectable [41], current survey data may not present the true picture of recent introductions. Therefore, we recommend that, in devising management strategies, the direct dispersal-based ranking is suitable for targeted short-term policies focusing on direct introduction (i.e. species in a location which pose a known risk), and the indirect dispersal-based ranking is a better guideline for long-term prevention strategies.

**Case study of higher-order species dispersal pathways**

In this section we present a case study of species introduction pathways through first-order and higher-order species flow networks. We focus on the port of Murmansk and compare the propagation pattern through the first-order node of Murmansk, the second-order node of Murmansk|Tromso, and Murmansk|Hammerfest. The first, second, and third step of propagation pathways for each port are visualized in Fig 6. Two key patterns emerge. First, a first-order model for simulating the propagation pattern results in several potential vulnerable ports that need to be targeted for management, even at the first step of propagation [compare Fig 6. (c1), Fig 6. (a1) and Fig 6. (b1)] corresponding to the first step of propagation for Murmansk, Murmansk|Hammerfest and Murmansk|Tromso respectively). These propagation pathways can become even more complex as we consider second and third step of species propagation. Considering the propagation through the second order ports, however, leaves far fewer targeted ports and highlights the few important pathways that should be targeted given each second-order node. As a result, using the first-order representation is not helpful when devising targeted control policies which require costly planning and resources.

Second, each second-order port has a unique propagation pattern resulting in different vulnerable ports at each step. For example, knowing that species were introduced from Hammerfest through Murmansk, the control policies should target Varandey and Longyearbyen to prevent the second step propagation (Fig 6. (a2)), while for species that were introduced from Tromso through Murmansk the prevention policies should target Varandey, Dudinka, Advent Bay, Torshavan, and Reykjavik (Fig 6. (b2)). SF-HON provides a more fine-grained view of species spread pathways based on previous destinations. Therefore, preliminary prevention strategies based on SF-HON can optimize the cost of management by accurate identification of the potential spread pathways based on species origin.

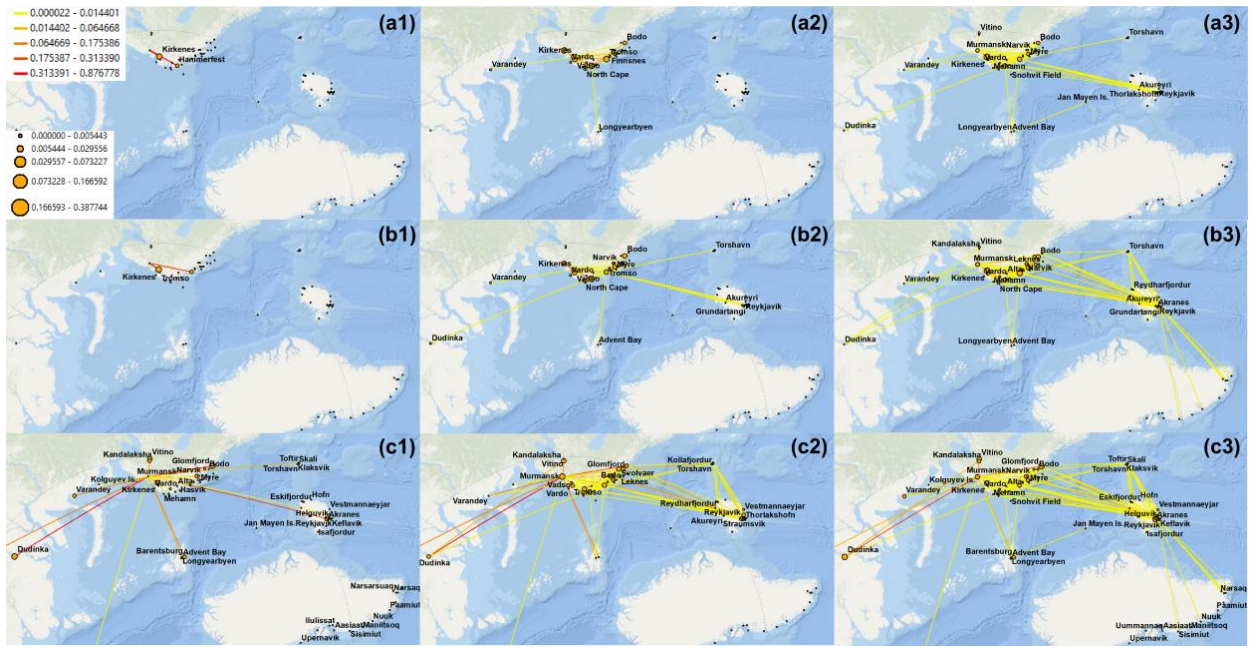

**Fig 6. Case study for higher-order introduction patterns of species.** Propagation steps through the second-order node of Murmansk|Hammerfest (a1), (a2), (a3). Propagation steps through the second-order node of Murmansk|Tromso (b1), (b2), (a3). Propagation steps through the first-order node of Murmansk (c1), (c2), (c3). The numbers in each case show the propagation step. (e.g. the first step of propagation through Murmansk is shown in (c1)). The size of the ports indicates the risk of species dispersal at each propagation step. The high-risk dispersal pathways are colored in red. In figures (c1), (c2), and (c3) the longest dispersal pathway (going outside the bottom border of the figures) connects Murmansk to Anadyr.

## Discussion

**Key observations and applicability to management**

Our work presents the most comprehensive assessment and projection available of ballast-mediated non-native species introduction and establishment into the Arctic, and the first attempt to integrate higher-order network approaches to analyze the within-Arctic dispersal risk. While future improvements in the data sets available and modeling methods used will build upon these results, the current analysis represents first guidance as to which ports might be prioritized in management efforts, given limited resources and the Arctic Council's goal of protecting the Arctic region from invasion [22]. The UN's International Maritime Organization's (IMO) 2004 International Convention for the Control and Management of Ships' Ballast Water and Sediments came into force in September 2017 and will provide a uniform international policy framework for reducing ballast-mediated introductions. Management mandated by the IMO convention could be augmented by geographically targeted efforts coordinated among Arctic Council member nations to protect the Arctic from threats revealed by our analyses.

Our analyses of the past 15 years' trend presents a first glimpse into the future risk posed to the pan-Arctic by ballast-mediated species introduction. Given current climate projections and the positive climate feedbacks on sea ice loss, our simple linear projection of current trends of shipping is no doubt wrong, and is more likely to be conservative than an overestimate. It indicates that shipping intensity (number of trips, average or aggregated ship capacities) and the average risk per introduction or diffusion pathway is likely to continue to increase. While the number of pathways may remain steady, the pathways are likely to be rerouted through Arctic shipping hubs such as Murmansk. Increased management at these emerging hubs could therefore have a disproportionately positive impact on risk of biological invasions across the entire Arctic.

Our analysis of species introduction pathways highlights the potential for the introduction of highly tolerant species from distant locations normally thought of as being disconnected from the Arctic, including Australia, South America, and Africa. It also reveals high-risk introduction pathways densely distributed in Northwestern Europe (the Murmansk—Narvik region). We further demonstrated how the same routes pose different levels of risk for different species, depending on their environmental tolerances. Thus, our analyses can be used at the general level, route-specific level or species-specific level (based on knowledge of a species' environmental tolerances).

Our analysis of ballast-mediated species dispersal within the Arctic leverages the higher-order network to capture the influence on species dispersal of path-dependent ship movement patterns. The higher-order network approach reveals port clusters between which only loose connects exist (e.g., a single route connects Alaskan ports to the other port clusters), highlighting opportunities to effectively prevent or slow species dispersal within the Arctic by improved management on a small number of inter-cluster routes. The higher-order network analysis also models indirect species dispersal, highlighting ports that might otherwise be ignored by management due to their weak connections to other ports, but which are still at risk through indirect dispersal in the long term. Our case study of species propagation in the Arctic ports highlights the advantages of using a higher-order network model for effective and accurate targeting of the potential vulnerable ports for early prevention strategies.

Overall the current results could aid in the development of effective Arctic invasive species management policies. As summarized above, our risk assessment framework identifies shipping routes where ballast water management will have the greatest impact on the overall threat of species invasion. Given limited resources and the especially challenging Arctic environment, information like we provide can help guide the placement of resources for surveillance, prevention, and other species management strategies. Place-based strategies that decision-makers could consider include: (1) prioritizing locations for surveillance efforts, including new eDNA and other genetic-based methods, to inform early detection and rapid response efforts [42],[43]; (2) choosing the location for on-shore ballast water treatment facilities for high-risk regions or ports that are becoming hubs (e.g., Murmansk and Narvik where many high risk pathways connect) [44]; (3) implementing strict management policies on the highest risk routes and ports, including routes through inter-cluster connections in the Arctic, and ports (e.g., Reykjavik) with a high potential of leading to further invasions; and (4) developing site-based and/or mobile equipment and protocols for rapid response control efforts when an incipient invasion is discovered.

**Benefits of the higher-order network approach**

Higher-order structures in networks influence clustering, ranking, anomaly detection, representation learning, and diffusion in networks [30],[36],[45],[46], suggesting that these processes are important for understanding patterns of invasion risk within shipping networks. In an attempt to capture higher-order movements in their shipping networks, Keller et al. [24] assumed a fifth-order for all shipping paths while Seebens et al. [6] included the entire trajectory of ship movements in the species invasion risk. Although assigning a fixed order of dependecy for the shipping trajectories or exhaustively including all possible higher-order dependecies obtained from the entire shipping trajectories may fit well for the historically observed data, they can overcomplicate the ship movement models by forcing higher-order patterns when first-order is sufficient, and risk losing the generality towards the future by overfitting the history. To preserve generality, it is necessary to learn which higher-order patterns are most likely to represent invasion risk, and which occur at frequencies low enough to represent low risk. Xu et al. [30] proposed a higher-order network construction algorithm that only keeps higher-order structures when they are significantly different than that of lower-orders, but that algorithm treats all ships in the global shipping network equally and cannot be readily adapted to species introduction risk modeling. Saebi et al. [31] extendted Xu et al.'s higher-order network algorithm [30] so that it respects the impact of ship type, ship size, trip duration and ballast water discharge patterns for more realistically modeling of species introduction. As a result, we apply the algorithm proposed in [31] to ensure that the networks presented here are a succinct reflection of the backbone of species introduction and dispersal pathways in the Arctic.

**Opportunities for future improvement and application**

We hope that our approach will be refined in the future in several directions. First, our framework can benefit from an improvement of the breadth of existing data set. For example, although the LLI data set we used is the most comprehensive data set available on global shipping, it has limited observations in the northern regions of Canada [1]–[4][47]. The NBIC ballast water data set, from which we parameterized the ballast water discharge model, is limited to the U.S. More comprehensive information about port types, port usage patterns and ships would allow for the implementation of machine learning approaches to modeling ballast water discharge and species introduction risks, producing more robust estimates of absolute risk.

Second, our framework can benefit from the real-time incorporation of new data that captures other aspects of rapidly changing conditions in the Arctic. Given melting Arctic sea ice and other warming temperatures, temperature and salinity in ports and other coastal environments may change rapidly, underscoring the need for the incorporation of more frequent and widespread environmental data. Future research could also couple ballast water discharge monitoring, species occurrence monitoring, and long-term monitoring of the introductions processes to complement existing data.

Third, our framework may benefit from more detailed biological models. For example, some of the key biological relationships—the likelihood of species establishment as a function of the volume of ballast discharge or discharge frequency or concentration of organisms in discharge—could be improved with more empirical research, although the impact of such changes could be incremental and uncertain. Thus, given the nature of the fast-changing and non-comprehensive data, our analyses do not attempt to compute the absolute magnitude of introductions risks. Rather we use available data to estimate relative risks, which we believe are far more robust than any estimates of the absolute risk.

Finally, our risk assessment and prediction framework can be extended in multiple ways. The absolute risk of invasion from biofouling is considered similar to that from ballast discharges [48], but is less studied and has not received policy attention like ballast-mediated introduction. However, when more data become available on biofouling patterns on different surfaces of different ship types, trip speed and detailed route trajectories (the satellite-based Automatic Identification System shows promise), our framework for assessing ballast water mediated introduction risk can be adapted to assess biofouling risk. Our framework can be paired with species monitoring efforts, to produce regions of interest to guide species monitoring, and validate our model framework using observed changes in species distributions.

## Methods

### Data sets and preprocessing

For ship movements, we utilized global ship movement data from the Lloyd's List Intelligence (LLI) for the years 1997, 1999, 2002, 2005, 2008, and 2012 (starting on May 1st of these years and ending on Apr 30th of the following years). This data set was organized by individual voyages, totaling 12,723,028 records across the six years. The data also included unique ship identifiers, ship type (150 categories), gross weight tonnage, dead weight tonnage, ship departure and arrival port and dates. Removing duplicate records yields 9,569,619 ship movements. The data was further subset to include only voyages to and between Arctic ports, herein defined as ports located within the boundary laid out by the Arctic Council's Conservation of Arctic Flora and Fauna working group [25], yielding 48,364 voyages through 3,902 introduction pathways from non-Arctic ports to Arctic ports, and 4,715 voyages through 1,269 dispersal pathways within the Arctic.

For ballast water discharge, we used the records collected from the U.S. National Ballast Water Information Clearinghouse (NBIC) [25] for the years 2004 to 2016 for ships completing foreign and domestic voyages in Alaska, totaling 4,926 records. This data set included information about vessel type (9 categories), ballast water discharge and gross weight tonnage. Records with missing information, zero discharges, or where recorded ballast water discharge exceeded recorded ballast water capacity were removed, leaving 1,280 valid records. Since ships sailing to and through the Arctic face unique climate conditions and have distinct patterns (frequencies and amounts) of ballast water update/discharge, we subset the NBIC data to include only voyages to and between Arctic ports, and used that subset in our estimates of ballast water discharge patterns.

For annual average water temperature and salinity conditions, we used the Global Ports Database [24] wherever possible, supplemented by records of surface water conditions closest to ports' locations from the World Ocean Atlas [28]. For ecoregion data, we used Marine Ecoregion of the World (MEOW) [26] and Freshwater Ecoregion of the World (FEOW) [27].

**Calculation of introduction risks**

The relative risk of invasive species introduction was first calculated for every ship movement in the Lloyd's data, then aggregated for every pathway. Inspired by the introduction risk equation from Seebens et al. [6], for a ship $s$ making the trip $t$ from port $i$ to port $j$, which took $\Delta_{i \to j}^{(t)}$ days and discharged $D_{i \to j}^{(t)}$ ballast water, the relative risk of introduction for this trip is:

$$P_{i \to j}^{(t)} = (1 - e^{-\lambda D_{i \to j}^{(t)}}) e^{-\mu \Delta_{i \to j}^{(t)}}$$

based on the premise that species have a higher probability of being transported through the trip if there was larger amount of ballast water discharge, or if the trip was short increasing the probability of species survival in the ballast. The duration $\Delta_{i \to j}^{(t)}$ of trip $t$ was taken from the Lloyd's data, and the daily species mortality rate $\mu = 0.02$ was chosen based on the work of Seebens et al. [6]. The species introduction potential per volume of discharge parameter $\lambda$ was given as $\lambda = 3.22 \times 10^{-6}$ based on Xu et al [16][11], so that $P_{i \to j}^{(t)}$ is 0.8 when ballast discharge volume is 500,000m³ and trip duration is zero. The volume of ballast water $D_{i \to j}^{(t)}$ translocated by trip $t$ made by a ship of type $k$ and gross weight tonnage $GWT$ is estimated using a modified version of the approach of Seebens et al. [6]:

$$D_{i \to j}^{(t)} = Z_k W_{GWT}$$

where $Z_k$ is the fraction of non-zero releases for ship type $k$, and $W_{GWT}$ is the estimated discharge in metric tons for a ship with gross weight tonnage $GWT$. The 150 ship types in the LLI data were mapped to the 9 types present in the NBIC data and given ship type $k$ from the 9 types in the NBIC data, the ballast water discharge frequency $Z_k$ is computed based on the NBIC data, yielding the mapping of $k \to Z_k$ (Supplementary Table S5). We estimated $W_{GWT}$, by removing the zero discharge records from the NBIC data, randomly splitting the data into training (70%) and testing (30%) data sets, and fitting a random forest regression to the training data in R. The random forest regression predicted $W_{GWT}$ as a function of ship type and gross weight tonnage and was validated using the testing set yielding an $R^2$ of 0.93 (Supplementary Fig. S1).

We computed the introduction risk $P_{i \to j}^{(t)}$ based on ship size, ship type, and trip duration. Assuming $P_{i \to j}^{(t)}$ are independent, then the aggregated probability of introduction for non-native species for a pathway $i \to j$ is:

$$P_{i \to j} = 1 - \prod_t (1 - P_{i \to j}^{(t)})$$

and the aggregated probability of introduction for a given target port $j$ is:

$$P_j = 1 - \prod_i (1 - P_{i \to j})$$

**Within-Arctic species dispersal higher-order network**

In the original work that proposed the higher-order network (HON) [30], the algorithm constructs the network based on a sole source of data (e.g., ship trajectories), which cannot readily fit the needs of modeling species dispersal that is a function of multiple factors. In this work, we use SF-HON, the extension of [30] proposed by Saebi et al. [31] since it can take an arbitrary number of data sources as input, can have customized aggregation functions, and can have customized thresholds.

To clarify the SF-HON approach, we present a side-by-side comparison of the original algorithm and the extended algorithm used in this work, illustrated in Supplementary Fig. S2, to highlight the key differences in network construction. The original HON algorithm takes a single source of event sequence (illustrated as ship trajectories) as the input, whereas the species-flow higher-order network (SF-HON) not only takes ship trajectories but also ship types, trip durations and ballast water discharge for every ship movement as the input. The influence per trip is trivial in HON, in that every ship movement is treated as having the same weight and counted as "one trip" from source port to target port; in SF-HON however, the influence of every trip can be different due to variations in trip duration, ballast discharge, ship type and so on, therefore we compute the risk of introduction $P_{i \to j}^{(t)}$ separately for every trip. Aggregating the influence through a given pathway in SF-HON is accomplished simply by counting the number of trips observed through the pathway; in SF-HON we instead take the joint probability assuming different trips are independent. Finally, as the algorithm needs a parameter "minimum support" as the terminating condition for higher-order rule extraction, in SF-HON the minimum support is a positive integer, such that pathways with less than the specified trips through them will be discarded. In SF-HON the minimum support is extended to probabilities, and pathways with aggregated probability of species introduction less than the specified threshold will be discarded.

For further analysis of Arctic dispersal pathways, we apply the clustering algorithm developed by Blondel et al. [49] to assign different colors to clusters in Fig. 5C. We use random walk with resets to calculate the ranking of ports based on species dispersal risk in Fig. 5C. While species dispersal dynamics can be approximated with random walks, realistically, we require the random walkers occasionally be relocated to a random Arctic port, so that (1) the random walkers will not be trapped in loops forever, and (2) random walkers have a small chance of following unobserved or new shipping paths. The idea of using "random walking with resets" to simulate multi-step species dispersal is inspired by the well-established PageRank algorithm [38], which was originally developed to model Web users' browsing behavior and rank Web pages' probabilities of being visited by a random user. Similarly, in the species dispersal context, a port's risk of receiving indirect species dispersal is the probability of being visited by a random walker flowing through the species dispersal network, denoted as the node size in Fig. 5C.

**Availability of materials and data**

The datasets generated during the current study are available in the following repository: https://github.com/xyjprc/SF-HON/

**Acknowledgements**

This paper is based on research supported by the U.S. National Science Foundation Award #1427157 (PIs: DM Lodge, NV Chawla, EK Grey), the Army Research Laboratory under Cooperative Agreement Number W911NF-09-2-0053 (PI: NV Chawla), the University of Notre Dame via a Jefferson Science Fellowship in the US Department of State (to DM Lodge), and University of Notre Dame Riley Center. For comments and suggestions, we thank the US Department of State Diplomacy Lab; Ann Meceda and Adrianna Muir of the US Department of State, and Peter Oppenheimer of US NOAA. We thank Kristy Deiner, James Corbett, Melinda Gormley, Amanda Leister, Anaïs Lacoursière-Roussel, and Kimberly Howland for discussions. We thank Kaitlin Jacobson and Savannah Wunderlich for literature survey and data analysis on the case studies, and other students in the Notre Dame Diplomacy Lab course (DM Lodge, instructor) for discussion.

**Competing interests**

The author(s) declare no competing interests.

**Author contributions**

All authors collectively conceived the research, designed the analyses, and interpreted results. M.S. and J.X. conducted the experiments. S.C. collected NBIC data and performed ballast water estimation. M.S., J.X., S.C., E.G., N.V.C., and D.M.L wrote the manuscript. M.S., J. X., and S.C. contributed equally to the manuscript. All authors reviewed the manuscript.